\documentclass[twocolumn,showpacs,prl,aps,amsmath,amssymb,
]{revtex4}       

\usepackage{graphicx}
\usepackage{dcolumn}
\usepackage{bm}
\usepackage{amssymb}

\begin{document}

\title{Multi-channel architecture for electronic quantum-Hall interferometry}

\author{Vittorio Giovannetti$^1$, Fabio Taddei$^1$, Diego Frustaglia$^{2}$, and
        Rosario Fazio$^{3,1}$}
\affiliation{$^1$NEST-CNR-INFM and Scuola Normale Superiore, I-56126 Pisa, Italy \\
	    $^2$ Departamento de F\'isica Aplicada II, 
	Universidad de Sevilla, E-41012 Sevilla, Spain\\
	$^3$International School for Advanced Studies (SISSA), I-34014 Trieste, Italy}

\date{\today}

\begin{abstract}
We propose a new architecture for implementing electronic interferometry in quantum Hall bars.
It exploits scattering among parallel edge channels. In contrast to previous developments, 
this one employs a simply-connected mesa admitting serial concatenation of building 
elements closer to optical analogues. 
Implementations of Mach-Zehnder and Hambury-Brown-Twiss interferometers are discussed 
together with new structures yet unexplored in quantum electronics.
\end{abstract}


\maketitle
{  Since many decades} interferometry {  has been} a fundamental tool to disclose the 
classical and quantum properties of light~\cite{MZ}. Nowadays optical interferometry 
can be considered at the heart of a new quantum-based  technology with 
applications in  metrology~\cite{METRO}, imaging~\cite{IMAGING}, and quantum 
information processing~\cite{OQC}. In the solid state world, controlled quantum 
interference experiments appeared more recently  when, thanks to the advances 
in fabrication, the wave-like nature of electrons could be tested in transport 
measurements. The observation of Aharonov-Bohm (AB) oscillations in the {  electric 
current~\cite{RING} and} the Landauer-B\"{u}ttiker formulation of 
quantum transport in terms of electronic transmission {  amplitudes}~\cite{lb} signaled 
the beginning of quantum electronic interferometry in solid state devices. Since 
then, there has been a continuous effort in studying interference effect in quantum 
transport~\cite{datta}. A recent breakthrough in {  electronic interferometry has been} 
the experimental realization of electronic Mach-Zehnder~\cite{Ji03,neder,litvin,preden,nedernaturep3} (MZ) 
and Hanbury-Brown-Twiss (HBT)~\cite{neder2} interferometers using edge states
in a quantum Hall bar. In these experiments electrons loop around an annular 
(Corbino-like) sample flowing along chiral edge channels which mimic the optical 
paths~\cite{edgestate}. In order to unfold the full potentiality of optical 
interferometry in the solid state realm an important additional ingredient
is needed: The ability to concatenate in series several MZ interferometers. {  This 
requirement, impossible to implement at present, leads us to develop a new 
interferometric architecture for edge states. This scheme opens up a wide range 
of new possibilities in electronic interferometry. As first examples  we discuss 
implementations of the MZ, HBT, and {\em interaction-free}~\cite{Jang} interferometers. 
Furthermore we show how to exploit our setup for characterizing the sources of dephasing 
in quantum Hall systems.}

\begin{figure}[t!]
\includegraphics[width=\columnwidth,clip]{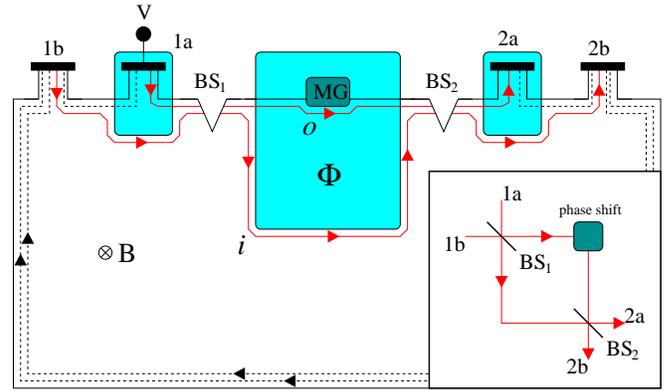}
\caption{(color online) Mach-Zehnder interferometer implementation. 
The shaded areas represent top gates which define regions of filling factor $\nu=1$.
In the rest of the sample instead we assume $\nu=2$.
Due to the presence of a strong magnetic field $B$ orthogonal to the bar surface,
electrons injected from the source contact $1$a propagate from left to right following 
two possible paths. Electrons are finally collected at drain contacts $2$a and $2$b, 
where currents are measured. The red lines represent the edge channels effectively taking 
part to the MZ interferometer. BS$_1$ and BS$_2$ are beam splitters, while the gate MG 
is used to vary the shape and length of the outer edge $o$ channel. Notice that an experimental 
implementation of the setup does not require to employ air-bridge elements. Inset: sketch 
of the optical counterpart of the MZ interferometer.}
\label{fig1}
\end{figure}

A good starting point to present our new architecture is to consider the MZ configuration.
In an optical MZ interferometer (inset Fig.~\ref{fig1}), a monochromatic beam from 
source $1$a  {  is} split into two beams by a beam splitter BS$_1$. {  The beams then 
propagate} along two different paths which recombine at a second {  beam splitter} BS$_2$, 
where interference occurs. The two outgoing beams are collected at detectors $2$a and $2$b.
In the absence of external noise, the beam intensity at the detectors exhibits oscillations 
as a function of the accumulated phase difference between the followed paths.
Our electronic implementation of the MZ interferometer is sketched in Fig.~\ref{fig1}. It 
consists of a 2DEG subject to a quantizing perpendicular magnetic field $B$ corresponding to 
a filling factor (number of occupied Landau levels) $\nu \equiv n_{\text{s}}h/eB=2$.
Four electronic contacts are present in the structure: A bias voltage $V$ is applied 
to $1$a, acting as a source, while the remaining contacts 1b, 2a and 2b are grounded.
The shadowed regions in the figure represent top gates which reduce the local electron density 
$n_{\text{s}}$ in such a way that the filling factor in the underneath regions is $\nu=1$. Such 
cross-gate technique, implemented e.g. in Refs.~\cite{wurtz,crossgate,haug}, is used to 
selectively address the two edge channels by introducing a spatial separation between them. 
In particular, the gate on  top of  $1$a  allows us to selectively populate only the outer edge 
${o}$ of the sample by preventing the inner channel ${i}$ to be subject to the bias voltage $V$.
Analogously, the gate on top of the contact 2a allows us to measure the current carried by 
the outer edge channel only. Finally the large top gate in the center of the setup induces 
a spatial separation between the two edge states. The area $A$ defined by the two paths 
encloses a magnetic flux $\Phi=BA$. It is important to notice that such an area can be substantially 
smaller as compared with other MZ realizations~\cite{Ji03,neder,litvin,preden,nedernaturep3,neder2}. {  In our 
proposed architecture} values of $A \sim 1$ $\mu$m$^2$ (corresponding to about $10^3$ 
flux quanta) are experimentally feasible with present technology. This is an improvement of 
almost two orders of magnitude with respect to conventional MZ setups that would arguably lead 
to a reduced effect of phase averaging due to area and/or flux fluctuations (as stated in 
Ref.~\cite{neder2}, where a visibility enhancement was ascribed to a size reduction with 
respect to previous implementations~\cite{Ji03}).

Beam splitter transformations among the edges ${o}$ and ${i}$ are introduced,
 as in Ref.~\cite{beenakker}, by
inducing elastic inter-channel scattering within the regions BS$_1$ and BS$_2$ of Fig.~\ref{fig1}.
{  This is admittedly the most delicate part of our proposal. There are
however two ways to implement it.
Inter-channel scattering can be obtained by an abrupt (non-adiabatic) variation in 
the confining potential such as the triangular-shaped protuberance shown in 
the figure. According to the calculations of Ref.~\cite{palacios}, edge channels mix 
coherently if the (potential defining) the protuberance shows spatial inhomogeneities 
on a scale smaller than the magnetic length $l_{\text{m}}=\sqrt{\hbar/eB}$. Such potential 
profiles can be engineered to give the desired scattering amplitude, for example, by 
the cleave-edge overgrowth technique~\cite{ceo}. Another possibility to have elastic
inter-channel scattering is to use high-spatial-resolution local probes as atomic force 
microscopy~\cite{woodside} or scanning gate microscopy~\cite{aoki}. In this way there
is the additional advantage that the scattering amplitudes can be tuned by means of 
an external voltage.}

As in Refs.~\cite{neder,neder2,Ji03,litvin,preden,nedernaturep3}, we assume the size of the structure to be much 
smaller than the equilibration length
at which spontaneous inter-channel mixing occurs~\cite{haug}. 
Moreover, apart from a small region where the BSs are implemented, 
the confining potential is 
assumed to be sufficiently smooth to prevent undesired inter-channel scattering.
Under these conditions, ${o}$ and ${i}$ represent two independent electronic modes of 
propagation which play the role of optical paths in a MZ interferometer.

\begin{figure}[t!]
\includegraphics[width=\columnwidth,clip]{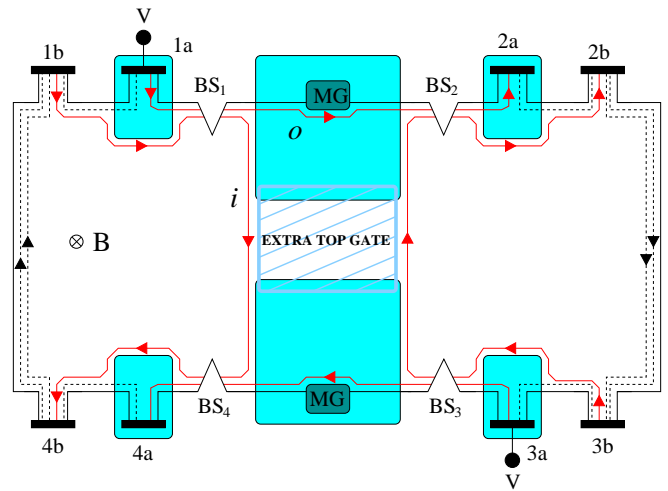}
\caption{(color online) Implementation of the  Hanbury-Brown-Twiss interferometer.}
\label{fig3}
\end{figure}

To have a first glimpse of the extreme versatility of {  this new} architecture we notice
that the setup of Fig.~\ref{fig1} can be turned easily into a HBT interferometer
--- see Fig.~\ref{fig3}. The resulting implementation is reminiscent  of the  one 
realized by  Neder {\em et al.} with  the traditional (non simply connected) 
mesa configuration~\cite{neder2}. In our case it has been obtained by introducing 
a further MZ interferometer in the bottom part of the mesa of Fig.~\ref{fig1}, 
allowing the central top gates to overlap (possibly with the help
of the extra top gate shown in the figure).
 
\begin{figure}[t!]
\includegraphics[width=\columnwidth,clip]{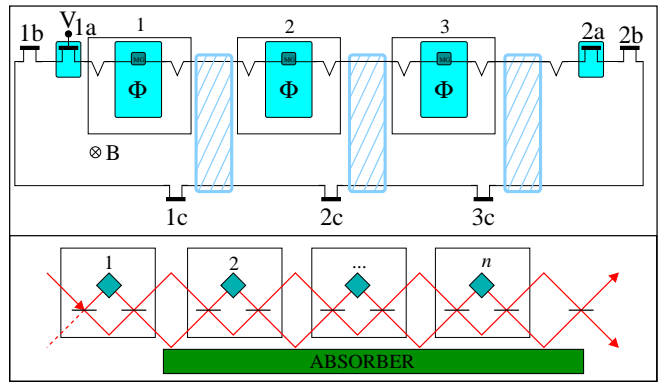}
\caption{(color online) {\em Interaction-free} interferometer.
The optical implementation is shown in the lower part of the graph. It consists
in a series of $n$ concatenated MZ interferometers characterized by a phases difference
$\phi = \pi/n$ among the two internal paths~\cite{Jang}.
One of the two emerging path from each of the
MZ impinges into an external absorber (the green box in the picture) which can be either
totally reflecting $(\eta =1)$  or totally absorbing $(\eta=0)$. 
Incoming photons deterministically end up either in the upper or the lower exit port depending 
on $\eta$. 
The electronic implementation
of this devices for $n=3$ is found in the upper part of the picture. 
Here the ``absorber'' is simulated by the grounded contacts $1c$, $2c$ and $3c$.
The absorption ($\eta=0$) case is simulated by switching on extra top gates (patterned areas 
in the figure) which put in contact the inner edge with $1c$, $2c$ and $3c$. 
  }
\label{fig1011}
\end{figure}

Additionally, very interesting devices with no counterpart in conventional edge state setups 
can be devised by fully exploiting the concatenability of our simply connected architecture.
A first example is sketched in Fig.~\ref{fig1011}. This is an electronic equivalent of 
the optical {\em interaction-free} interferometer of Ref.~\cite{Jang} 
(see {  the} caption of Fig.~\ref{fig1011} for a brief description of its working principles). 
With our architecture we can reproduce it by properly concatenating a series of MZs of Fig.~\ref{fig1}. 

\begin{figure}[t!]
\includegraphics[width=\columnwidth,clip]{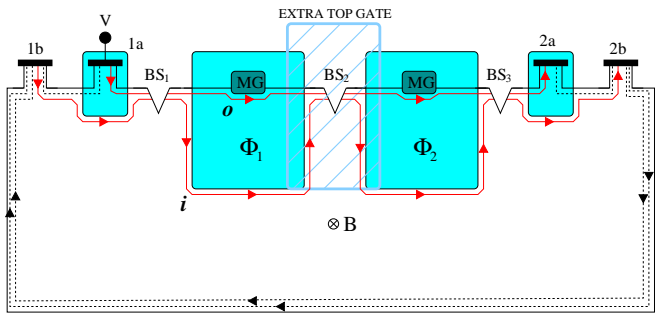}
\caption{(color online) Example of two MZ interferometers 
 concatenated. The two large shaded areas are are characterized respectively 
by AB fluxes $\Phi_{1}$ and
$\Phi_2$ and path-length differences $\Delta L_{1,2}$. 
The patterned 
area represents an auxiliary top gate which can be inserted to bypass
the second BS, converting the whole setup to a the single MZ interferometer
of Fig.~\ref{fig1}. 
}\label{fig2}
\end{figure}

\begin{figure}[t!]
\includegraphics[width=\columnwidth,clip]{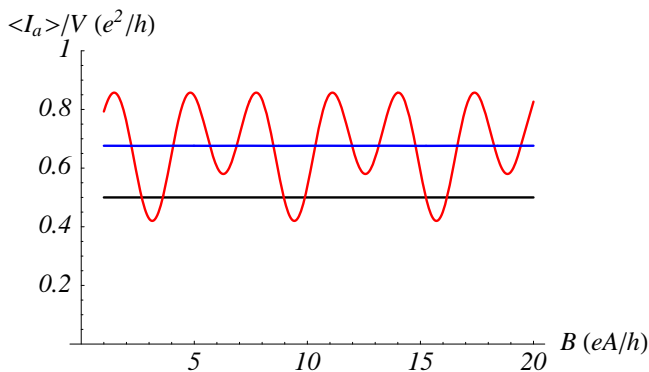}
\caption{(color online) Linear conductance $G_a=\langle I_{\text{a}}\rangle/V$ as a function of 
magnetic field $B$ for different values of voltage $V$, with $D_1=0.8$, $D_2=0.9$, 
$R_{3\text{u}}=0.4$, $T_{3\text{u}}=0.6$, $\Delta L_1/\Delta L_2-1=10^{-4}$.
Red (blue) curve is relative to a small (large) voltage, while the black curve
is the completely incoherent case (i.e. $D_{1,2}=0$).}
\label{graph}
\end{figure}

A further interesting application is found in the characterization of dephasing in quantum Hall systems 
{  which is recently attracting a lot of 
interest~\cite{Ji03,neder,litvin,chung2,preden,nedernaturep3,marquardt,NEDERNJP}. }
Consider first the setup of Fig.~\ref{fig1}. The transmission 
probability $T_{\text{a}}(E)$ from terminal 1a to 2a,  reads
\begin{eqnarray}
\label{Ta}
T_{\text{a}}(E) &=&
T_{1 {o}} T_{2 {o}}+R_{1 {o}} R_{2 {i}} \\ 
&+& 
2\sqrt{T_{1 {o}} T_{2{o}} R_{1{o}} R_{2 {i}}} 
\cos[\Delta\phi(E)] \;,\nonumber
\end{eqnarray}
where for $\alpha={o,i}$ and $j=1,2$,  $T_{j{\alpha}}$ and 
$R_{j{\alpha}}=1-T_{j{\alpha}}$ are, respectively, the transmission 
and the reflection probabilities of the beam splitter BS$_j$.
A similar expression can be obtained for $T_{\rm b}$ (transmission from 1a to 2b). 
The last contribution in Eq.~(\ref{Ta}) is an interference term leading to current 
oscillations at the contact $2\text{a}$. It accounts for the phase difference $\Delta\phi$ 
associated to the two possible paths the electrons can choose in their propagation. Apart 
from an irrelevant constant term, in the absence of external noise it can be expressed as
\begin{eqnarray}
\Delta\phi (E)= \phi_{\text{D}}(E) + \phi_{\text{AB}}\;. 
\label{phase}
\end{eqnarray} 
The first term is a dynamical contribution given by~\cite{chung2} $\phi_{\text{D}}(E) =E 
\Delta L/\hbar v_{\text{D}}$, where $\Delta L$ is length difference between the paths (for 
simplicity we assume the channels ${o}$ and ${i}$ to have identical drift velocities). The second term in 
Eq.~(\ref{phase}) is the {  AB contribution $\phi_{AB} = 2\pi e\Phi/h$.} Both the dynamical 
and the AB contributions can be varied in our setup by modifying the shape of the outer path 
by means of the local gate MG of Fig.~\ref{fig1}. {  Decoherence can be} 
described by adding an extra term $\varphi$ in Eq.(\ref{phase}) which accounts for  possible 
phase fluctuations. These may originate either from long time oscillations of locally trapped 
impurities or thermal fluctuations of the edge-state local density.
Decoherence eventually leads to the suppression of the interference term in $T_{\text{a,b}}(E)$, 
Eq.~(\ref{Ta}), {  thus} reducing the visibility of the oscillations induced by the modulation 
{  of the gate}
MG in the output currents~\cite{lb} 
$I_\text{a,b} \equiv (e/h) \int dE  [f(E-eV) - f(E)] T_{\text{a,b}}(E)$. 
Notably the visibility can be suppressed even in the absence of decoherence. 
This is due to the energy dependence of the phase, giving rise to a phase-averaging 
of the current $I_\text{a,b}$ when integrating over a large energy window~\cite{chung2}.
The visibility decrease has been intensively investigated
in these systems~\cite{Ji03,neder,litvin,chung2,preden,marquardt,NEDERNJP}
trying to identify its sources 
by means of shot noise measurement~\cite{Ji03}. {  No information} can be 
obtained just from the average current by using a MZ single interferometer. 
{  Our architecture makes possible to discriminate between phase-averaging 
and decoherence mechanisms directly in the measurement of the average current by 
concatenating two MZ interferometers.}
%
%
The setup is shown in Fig.~\ref{fig2}.  For the sake of simplicity we assume 
$T_{1,2\alpha}=R_{1,2\alpha}=1/2$. The transmission between 1a and 2a for this device reads
\begin{eqnarray}\label{tr2MZ}
T_{\text{a}}(E)&=&
1/2+(R_{3{o}}-T_{3{o}})\cos[\Delta\phi_1(E) + \varphi_1]/2 \\
&+&\sqrt{R_{3{o}} T_{3{o}}} \sin[\Delta\phi_1(E)+ \varphi_1] 
\sin[\Delta\phi_2(E) + \varphi_2]\;, \nonumber 
\end{eqnarray}
with $\Delta \phi_{1,2}(E)$ defined, as in Eq.~(\ref{phase}), in terms of
the parameters $\Delta L_{1,2}$ and $\Phi_{1,2}$ associated with the two large gated areas of 
Fig.~\ref{fig2}. The $\varphi_{1,2}$ account for corresponding noise fluctuations.
In the linear-response regime at zero temperature, contact 2a receives an output current
$I_{\text{a}}= (e^2 V/2h)[1 + (R_{3{o}}- T_{3{o}})\; \kappa_1 + 2\sqrt{R_{3{o}} T_{3{o}}}\; \kappa_2]$
with 
\begin{eqnarray}\nonumber
\kappa_1 &=& \int_0^{eV} \frac{dE}{eV}  \cos[\Delta \phi_1(E) +\varphi_1],
\\
\kappa_2 &=& \int_0^{eV} \frac{dE}{eV}  \sin[\Delta \phi_1(E) + \varphi_1] \sin 
[\Delta \phi_2(E) + \varphi_2 ] \;. \nonumber 
\end{eqnarray}
Regarding decoherence, we treat it in a 
phenomenological fashion by defining a zero-temperature distribution of phase fluctuations 
$P(\varphi_1,\varphi_2)$,
such that the average current reads $\langle I_{\text{a}}\rangle = \int d\varphi_1 
d\varphi_2\; P(\varphi_1,\varphi_2) I_{\text{a}}$. In the uncorrelated case [i.e. $P(\varphi_1,
\varphi_2)=P_1(\varphi_1) P_2(\varphi_2)$] with Gaussian phase-fluctuations 
(of width $\sigma_{1,2}$) 
we find
\begin{eqnarray}
\langle I_{\text{a}}\rangle = \frac{e^2V}{2h}[1+(R_{3{o}}- T_{3{o}}) \tilde{\kappa}_1 D_1
 +
2\sqrt{R_{3{o}} T_{3{o}}}\tilde{\kappa}_2 D_1 D_2], \nonumber 
\end{eqnarray}
where $D_{1,2}\equiv\exp[-\sigma_{1,2}^2/2]$, $\tilde{\kappa}_1 \equiv \kappa_1(\varphi_1=0)$ 
and $\tilde{\kappa}_2 \equiv \kappa_2(\varphi_2=0)$. We see that $\langle I_{\text{a}}\rangle$ 
has two interference terms proportional to $\tilde{\kappa}_1$ and $\tilde{\kappa}_2$, 
respectively. The interference terms vanish {\it only} in the presence of full decoherence ($D_{1,2}=0$):
Strong phase averaging (large voltages) 
reduces $\tilde{\kappa}_1$ to zero, but geometrical correlations between $\Delta\phi_1$ and $\Delta\phi_2$  
can preserve $\tilde{\kappa}_2$ from that 
(for instance, the case $\Delta L_1=\Delta L_2$ and $\Phi_{1}=\Phi_{2}$ yields $\tilde{\kappa}_2\simeq 1/2$).
{  This is strikingly different from the results of a single MZ interferometer, where 
complete phase averaging leads to the suppression of any interference term in the current 
and hence one has to resort to shot noise measurements.} 
For illustration, we provide an example in Fig.~\ref{graph}. There we plot the linear conductance 
$G_a \equiv \langle I_{\text{a}}\rangle/V$ 
as a function of the magnetic field for small voltages $eV\ll \hbar v_D/\Delta L_1$ (red curve) and for 
large voltages $eV\gg \hbar v_D/\Delta L_1$ (blue curve) in the presence of a small decoherence
($D_{1,2} \simeq 1$). Oscillations are suppressed as voltage increases due to the voltage dependence 
of $\tilde{\kappa}_{1,2}$. For large voltages (blue curve), the conductance converges to a constant 
value $G_a = \frac{e^2}{2h}[1+D_1D_2\sqrt{R_{3{o}} T_{3{o}}}]$ depending only on decoherence
through $D_{1,2}$.
In the case of strong decoherence $G_a$ takes the universal value $e^2/2h$ (black curve) 
showing no voltage dependence.
A similar analysis was presented in Ref.~\cite{Ji03}.  In that case, shot-noise 
measurements were needed. {  It is also worth noticing} that the configuration of 
Fig.~\ref{fig2} can be used {  to explore possible spatial correlations} between the 
fluctuations in the two different MZ interferometers.

The architecture presented in this paper, once realized experimentally, may open 
up a way to an entire new class of electronic interferometry. We gave three examples,
all based on the concatenation of several MZ interferometers. This proposal can be 
easily generalized to filling factors higher than $2$, which would allow the 
implementation of complex multi-mode interferometry. This, along with the ability 
of multiple concatenation of interferometers, could yield prototypical 
implementation of simple linear-optics-like quantum computing~\cite{OQC} devices,
or be relevant in revealing non-Abelian statistics 
in the fractional Quantum Hall regime~\cite{law}.
Moreover, by properly tuning the BS transparencies, the setup of Fig.~\ref{fig1} yields 
a edge-channel swapper. Alternatively, it can be employed to prepare controlled 
superpositions of the two outgoing edge channels. 

We thank  M. Heiblum, F. Marquardt,  V. Piazza and S. Roddaro
 for comments and discussions.
We acknowledge financial support from the EU funded NanoSciERA ``NanoFridge'' 
and RTNNANO projects, and the ``Ram\'on y Cajal'' program of 
the Spanish Ministry of Education and Science.

\end{document}